\documentclass[epj,nopacs]{svjour}

\usepackage{graphicx}
\usepackage{multirow}
\usepackage{todonotes}
\usepackage{nicefrac}
\usepackage{amssymb}

\renewcommand{\vec}[1]{\mathbf{#1}}
\def\openone{\hbox{\upshape \small1\kern-3.3pt\normalsize1}}
\newcommand{\ellf}{\overline{\ell}_4}
\newcommand{\eqref}[1]{(\ref{#1})}

\begin{document}

\title{The scalar radius of the pion from Lattice QCD
       in the continuum limit}

\author{Vera G\"ulpers\inst{1} \and
        Georg von Hippel\inst{1} \and
        Hartmut Wittig\inst{1,2}}

\institute{PRISMA Cluster of Excellence and Institut f\"ur Kernphysik,
Johannes Gutenberg-Universit\"at Mainz, 55099 Mainz, Germany \and
Helmholtz Institute Mainz, Johannes Gutenberg-Universit\"at 
Mainz, 55099 Mainz, Germany}

\abstract{
We extend our study
\cite{Gulpers:2013uca}
of the pion scalar radius in two-flavour lattice QCD
to include two additional lattice spacings as well as lighter pion masses,
enabling us to perform a combined chiral and continuum extrapolation.
We find discretisation artefacts to be small for the radius,
and confirm the importance
of the disconnected diagrams in reproducing the correct chiral behaviour.
Our final result for the scalar radius of the pion at the physical point is
$\left\langle r^2\right\rangle^\pi_{_{\rm S}}=0.600\pm0.052~\textnormal{fm}^2$,
corresponding to a value of $\ellf=4.54\pm0.30$ for the low-energy constant
$\ellf$ of NLO chiral perturbation theory.
}

\PACS{
{12.38.Gc}{Lattice QCD simulations} \and
{14.40.Aq}{Pi, K and eta meson properties}
}

\maketitle

\section{Introduction}

Hadronic form factors carry crucial information about the internal structure
of hadrons as bound states of quarks and gluons in Quantum Chromodynamics (QCD).
Since the dynamics underlying this structure is deeply non-per\-tur\-ba\-tive,
the only known first-principles approach that allows for the extraction of form
factors from QCD with full control of systematic errors is the use of lattice
simulations.

The scalar form factor of the pion, defined as
\begin{equation}
  F^\pi_{_{\rm S}}\left(Q^2\right) \equiv
  \left<\pi^+\left(p_f\right)\right|\,m_{\rm d}\overline{d}d 
    +m_{\rm u}\overline{u}u\,\left|\pi^+\left(p_i\right)\right>
\label{eq:defff}
\end{equation}
with the four-momentum transfer
\begin{equation}
 Q^2=-q^2=-(p_f-p_i)^2\,,
\end{equation}
is not directly accessible to experiment for lack of a suitable
low-energy probe. However, the associated scalar radius
\begin{equation}
 \left\langle r^2\right\rangle^\pi_{_{\rm S}} = -\frac{6}{F^\pi_{_{\rm S}}(0)}
                  \frac{\partial F^\pi_{_{\rm S}}(Q^2)}{\partial
Q^2}\Big|_{Q^2=0}
\label{eq:scalarrdef}
\end{equation}
can be related to the experimentally measurable cross section for $\pi\pi$
scattering
\cite{Colangelo:2001df}
using chiral perturbation theory ($\chi$PT).
A notable feature of the scalar radius is that in $\chi$PT at NLO
it depends only on a single low-energy constant, $\ellf$, through
\cite{Gasser:1983yg,Gasser:1983kx}
\begin{equation}
\left\langle r^2\right\rangle^\pi_{_{\rm S}} = -\frac{1}{(4\pi F)^2}
\,\frac{13}{2}
+ \frac{6}{(4\pi F)^2}\left[\overline{\ell}_4
+ \ln\left(\frac{m_{\pi,\rm{phys}}^2}{m_\pi^2}\right)\right]
\label{eq:scalarrchipt}
\end{equation}
where the pion decay constant is $F=92.2$~MeV
\cite{PDG:2014}.
\begin{figure*}
\centering
\includegraphics[width=0.7\textwidth,keepaspectratio=]{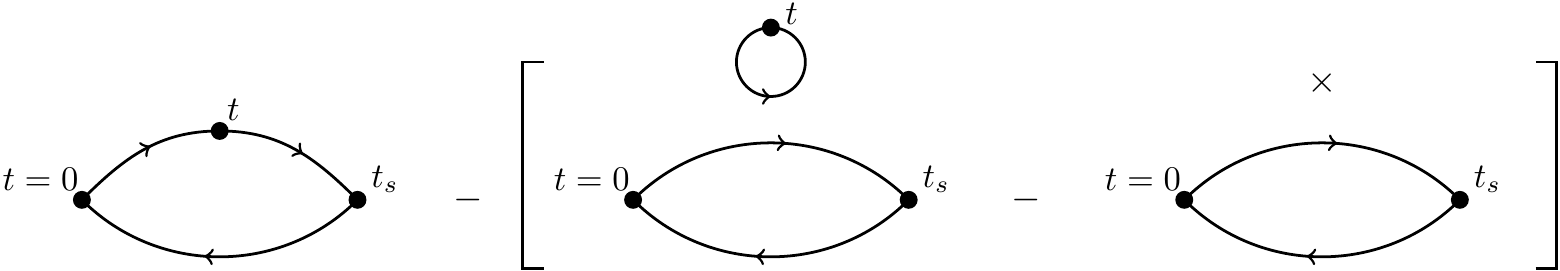}
\caption{Quark flow diagrams for the connected and disconnected contributions
         to the scalar form factor of the pion. The cross represents the
         vacuum expectation value $\langle\overline{\psi}\psi\rangle$,
         which for the
         Wilson fermion discretisation used here contains a power-law divergent
         additive renormalisation.}
\label{fig:diagrams}
\end{figure*}

In lattice QCD, the scalar form factor of the pion is derived from a
three-point function which receives contributions from both
quark-connected and quark-disconnected diagrams
(cf. figure~\ref{fig:diagrams}).
Due to the large numerical cost disconnected diagrams have often been neglected
in lattice calculations. However, arguments based on partially quenched
$\chi$PT 
\cite{Juttner:2011ur} 
indicate that quark-disconnected contributions to the
pion scalar radius could be sizeable.
Our earlier work 
\cite{Gulpers:2013uca}, 
using lattice data at a single value of the lattice spacing and relatively
large pion masses, has largely confirmed the prediction of $\chi$PT. In the
present work we extend our study significantly, by including two additional
lattice spacings, as well as lower pion masses. This allows for a combined
chiral and continuum extrapolation of our results. In this way we obtain
improved estimates for the pion scalar radius and the low-energy constant
$\ellf$, including a full assessment of systematic errors.

\section{Methods and Results}

\subsection{Ensembles used}

In our calculation of the scalar radius we use $N_f=2$ dynamical flavours of
non-perturbatively $O(a)$-improved Wilson fermions. The gauge ensembles have
been generated by the CLS initiative and are listed in table
\ref{tab:scalarffensembles}.

\begin{table}[h]
\centering
 \begin{tabular}{ccccccccccccccc}
\hline
 Label & $a [\textnormal{fm}]$ & lattice & $m_\pi [\textnormal{MeV}]$ &
$m_\pi L$ & $N_{\rm{cfg}}$\\
\hline\hline
 A3 & $0.079$ & $96\times48^3$ & 473  & 6.1 & $133$\\
 A4 & $0.079$ & $96\times48^3$ & 363  & 4.7 & $200$\\
 A5 & $0.079$ & $96\times48^3$ & 312  & 4.0 & $250$\\
 B6 & $0.079$ & $96\times48^3$ & 267  & 5.1 & $159$\\
\hline
 E3 & $0.063$  & $64\times32^3$ & 650 & 6.6 & $156$\\
 E4 & $0.063$  & $64\times32^3$ & 605 & 6.2 & $162$\\
 E5 & $0.063$  & $64\times32^3$ & 456 & 4.7 & $1000$\\
 F6 & $0.063$  & $96\times48^3$ & 325 & 5.0 & $300$\\
 F7 & $0.063$  & $96\times48^3$ & 277 & 4.3 & $351$\\
 G8 & $0.063$  & $128\times64^3$ & 193 & 4.0 & $348$\\
\hline
 N5 & $0.050$  & $96\times48^3$ & 430 & 5.2 & $477$ \\
 N6 & $0.050$  & $96\times48^3$ & 332 & 4.1 & $946$\\
 O7 & $0.050$  & $128\times64^3$ & 261 & 4.2 & $490$\\
\hline
 \end{tabular}
\caption{Overview of the CLS
	 ensembles that have been used for the calculation of the scalar
	 form factor of the pion. The lattice spacing
         given was determined using the $\Omega$ baryon mass
         \cite{Capitani:2011fg}.
         Note that all ensembles fulfil $m_\pi L\geq4$.}
\label{tab:scalarffensembles}
\end{table}

Compared to our previous work \cite{Gulpers:2013uca} we have included two more
lattice spacings, a coarser one, $a=0.079$~fm (labels A and B), and a finer one,
$a=0.050$~fm (labels N and O). For the intermediate lattice spacing,
$a=0.063$~fm, the G8 ensemble with a pion mass of $193$~MeV has been included
in our calculation. The use of different lattice spacings allows for performing
a continuum extrapolation of our results for the scalar radius besides the
chiral extrapolation. In general one also has to extrapolate to infinite volume,
however, all our ensembles fulfil $m_\pi L\geq4$ and thus we expect finite
volume effects to be negligible.

\subsection{Form factor calculation} 

We explicitly calculate the connected and the disconnected contribution to the
matrix element $\left<\pi\right|\overline{q}q \left|\pi\right>$ for three
different momentum transfers and
determine the scalar pion form factor. The three momentum transfers are chosen
such that the pion has momentum $|\vec{p}_i|=0$ at the source and momentum
$|\vec{p}_f|=0$, $|\vec{p}_f|=\nicefrac{2\pi}{L}$ or
$|\vec{p}_f|=\sqrt{2}\,\,\nicefrac{2\pi}{L}$ at the sink.

The quark loop
required for the disconnected part is estimated using three stochastic sources
per timeslice $t$ and a generalised Hopping Parameter Expansion to 6th order.
Further details can be found in \cite{Gulpers:2013uca}.

We use ratios \cite{Boyle:2007wg} of three- and two-point functions to extract
the desired scalar matrix element from our data. For the connected contribution
we use 
\begin{equation}
 R_1(t,t_s,\vec{p}_i,\vec{p}_f) = 
\sqrt{\frac{C_{3}(t,t_s,\vec{p}_i,\vec{p}_f)
 C_{3}(t,t_s,\vec{p}_f,\vec{p}_i)}
{C_{2}(t_s,\vec{p}_i)C_{2}(t_s,\vec{p}_f)}}\,,
\label{eq:R1}
\end{equation}
where the pion sink is located on $t_s$ and the scalar operator is inserted at
time $t$. The pion source is placed at $t=0$ for simplicity. The required
two-point and connected three-point functions have been calculated with 
Gaussian smearing
\cite{Gusken:1989ad,Alexandrou:1990dq,Allton:1993wc} 
applied to the source.

For the disconnected contribution we use smeared-smeared two-point correlation
functions and the ratio
\begin{eqnarray}
 R_3(t,t_s,\vec{p}_i,\vec{p}_f) &=&
\frac{C_{3}(t,t_s,\vec{p}_i,\vec{p}_f)}
{C_{2}(t_s,\vec{p}_f)} \\
&& \times \sqrt{\frac{C_{2} 
(t_s, \vec{p}_f)
C_{2}(t,\vec{p}_f)C_{2}((t_s-t),\vec{p}_i)}
{C_{2}(t_s,\vec{p}_i)C_{2}(t,\vec{p}_i)
C_{2}((t_s-t),\vec{p}_f) } }\,.\nonumber
\label{eq:R3}
\end{eqnarray}
To increase the statistics for the disconnected contribution we average over
four different pion source positions.

In our previous work \cite{Gulpers:2013uca} we have discussed the remaining
time-dependences of these ratios due to periodic boundary conditions and the
backward propagating pion in the two-point function. We take these dependences
into account by dividing the ratios $R_1$ and $R_3$ by the
appropriate time-dependent factors. Furthermore, we have shown significant
excited state contributions for small source-sink separations of
$t_s<1.5$~fm. This has been confirmed on our extended set of ensembles. Thus,
for all ensembles we only use data with $t_s>1.5$~fm.
Further details on the extraction of the scalar form factor can be found in
\cite{Gulpers:2013uca}.

\begin{figure}[h]
\centering
\includegraphics[width=0.5\textwidth]{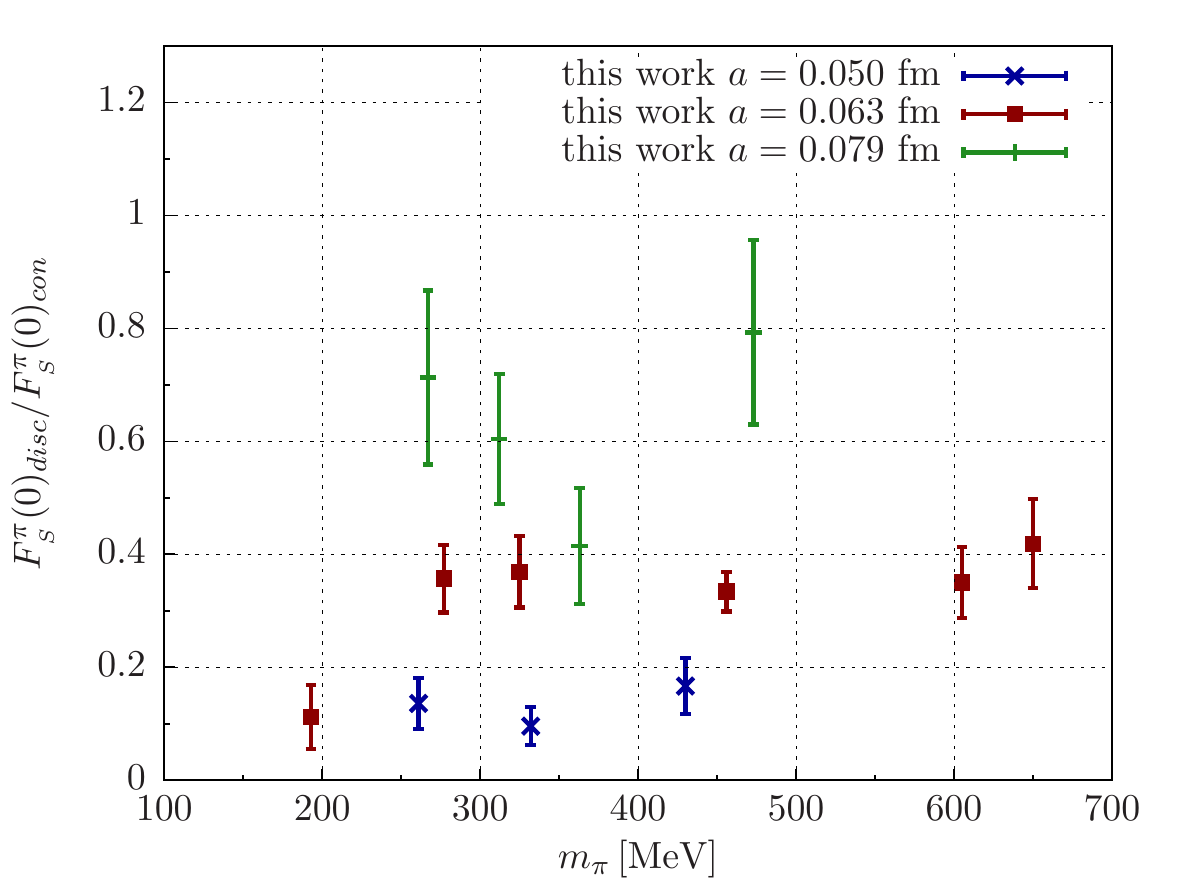}
\caption{The disconnected contribution
         $F^\pi_{_{\rm S}}(0)_{\rm disc}$ at $Q^2=0$
         divided by the corresponding connected contribution.
         Blue, red and green points denote different values of the
         lattice spacing $a=0.050$~fm, $0.063$~fm and $0.079$~fm,
         respectively.}
\label{fig:comp_disc_con}
\end{figure}

For all ensembles we observe a significant contribution of the disconnected
diagram to the scalar form factor. However, we find the relative
disconnected contribution to the form factor to be strongly dependent on
the lattice spacing. Figure \ref{fig:comp_disc_con} shows the disconnected
contribution at vanishing momentum transfer divided by the corresponding
connected contribution $F^\pi_{_{\rm S}}(0)_{\rm disc} /
F^\pi_{_{\rm S}}(0)_{\rm con}$ plotted against the pion mass.
Different colours denote different values of the lattice spacing.
Except for the lightest ensemble (G8), which might simply be an outlier, we
find no significant dependence on the pion mass. However, the relative
contribution of the disconnected diagram has a very pronounced dependence on
the lattice spacing. The smaller the lattice spacing, the smaller is the
relative disconnected contribution
$F^\pi_{_{\rm S}}(0)_{\rm disc}/F^\pi_{_{\rm
S}}(0)_{\rm con}$ to the form factor.

\subsection{The scalar radius}

 The scalar
radius of the pion is determined from the $Q^2$-dependence of the form factor at
vanishing momentum transfer. To estimate the scalar radius we use a linear
parametrisation, i.e.
\begin{equation}
 F^\pi_{_{\rm S}}(Q^2) = F^\pi_{_{\rm S}}(0)\left(1 -
\frac{1}{6}\left\langle r^2\right\rangle^\pi_{_{\rm S}}Q^2 +
\mathcal{O}(Q^4)\right)\,.
\label{eq:linparam}
\end{equation}
To determine the scalar radius on a given ensemble we fit a
function of the form \eqref{eq:linparam} to our form factor results at the
three different momentum transfers.
Note, that the absolute normalization of the form factor is not known, since
the multiplicative renormalisation $Z_s$ has not been determined. However, in
the calculation of the scalar radius $Z_s$ drops out.
For the lightest ensemble G8, the ratios for
the third momentum transfer were too noisy to obtain a useful signal. Thus, we
have to resort to match a linear function to two $Q^2$ values only for this
ensemble. For all other ensembles where data for three different momentum
transfers is available we find that the differences between using two or three
momenta is not significant.

\begin{figure}[h]
\centering
\includegraphics[width=0.5\textwidth]{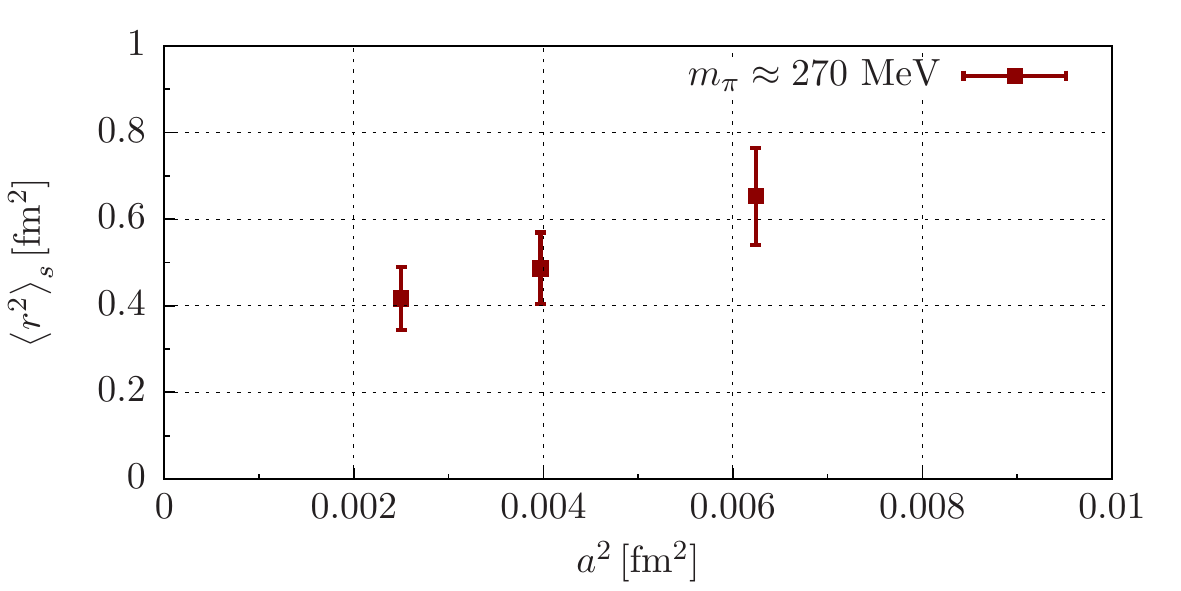}\\
\includegraphics[width=0.5\textwidth]{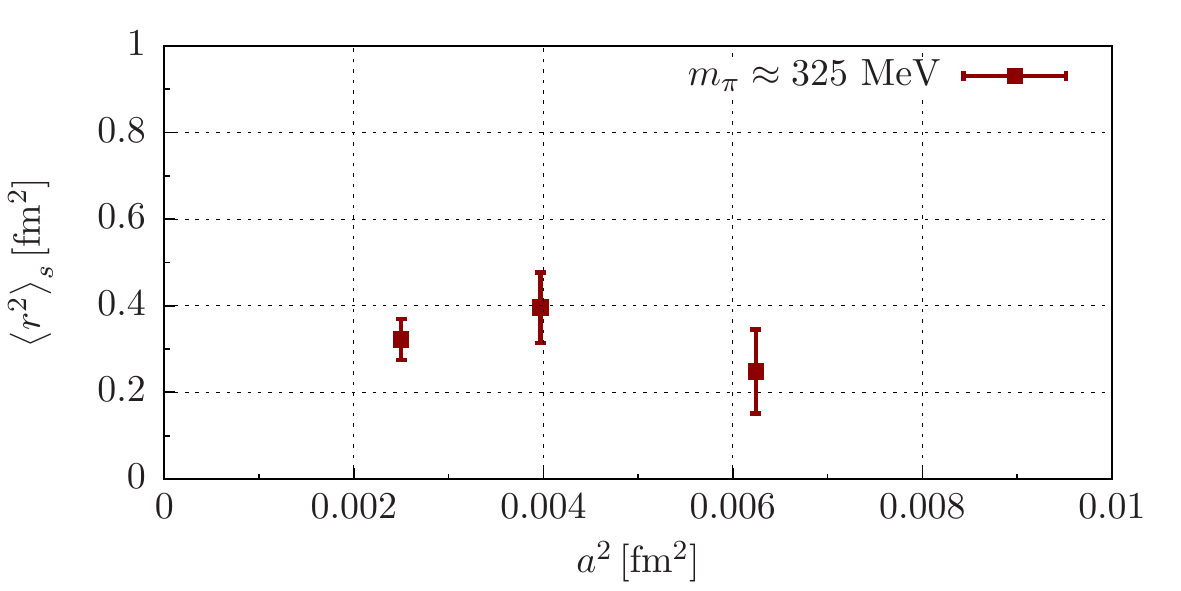}
\caption{The scalar radius at fixed pion mass plotted against the squared
         lattice spacing $a^2$. The upper panel shows results for a pion
         mass of $m_\pi\approx270$~MeV, the lower panel for
         $m_\pi\approx325$~MeV.}
\label{fig:radiusatfixedmass}
\end{figure}

To investigate a possible dependence of the scalar radius on the lattice spacing
we compare the results from ensembles with different lattice spacings at
roughly the same pion mass. Within the CLS ensembles we have two sets of such
ensembles (cf. table \ref{tab:scalarffensembles}): $m_\pi\approx270$~MeV (B6,
F7, O7) and $m_\pi\approx325$~MeV (A5, F6, N6). In
figure~\ref{fig:radiusatfixedmass} our results for the scalar radius on these
ensembles are plotted against the squared lattice spacing $a^2$. The upper
panel shows results for a pion mass of $m_\pi\approx270$~MeV the lower panel
for $m_\pi\approx325$~MeV.

While for the smaller pion mass one can see a trend in the scalar radius
for increasing lattice spacing, we do not observe such a trend for
$m_\pi=325$~MeV. Although we do not find a significant dependence on the
lattice spacing, we will include a term linear in $a^2$ for
performing a combined extrapolation of the scalar radius to the physical point.

\section{Chiral and continuum extrapolation}

The use of gauge ensembles with different pion masses and lattice spacings
allows for an extrapolation to the physical point $m_\pi\rightarrow
m_{\pi,{\rm phys}}$ and $a\rightarrow0$. For the dependence on the
pion mass we use the expression \eqref{eq:scalarrchipt} from NLO
$\chi$PT. Since we use an $O(a)$-improved action and because no operators that
mix with
$\overline{q}q$ at $O(a)$ exist, we expect lattice artefacts
to be of order $a^2$. Thus, we include a term $\propto a^2$ in a combined fit
of the scalar radius for all gauge ensembles (cf. table
\ref{tab:scalarffensembles}):
\begin{equation}
\left\langle r^2\right\rangle^\pi_{_{\rm S}} = -\frac{1}{(4\pi F)^2}
\,\frac{13}{2}
+ \frac{6}{(4\pi F)^2}\left[\overline{\ell}_4
+ \ln\left(\frac{m_{\pi,\rm{phys}}^2}{m_\pi^2}\right)\right] + b\,a^2\,.
\label{eq:scalarrchiptcont}
\end{equation}
The fit function \eqref{eq:scalarrchiptcont} has two fit parameters, the
low-energy constant $\ellf$ and the coefficient $b$ in the term which depends on
the lattice spacing. When fitting our data for the scalar radius to the NLO
expression \eqref{eq:scalarrchiptcont} we apply a cut in the pion mass by
imposing $m_\pi<500$~MeV which excludes the ensembles E3 and E4.

Additionally, we have repeated the fit using all data points (i.e.\ no mass cut)
and a more aggressive mass cut $m_\pi<335$~MeV. We find that the fit result is
quite insensitive to mass cuts, and the fits produce consistent results with an
acceptable $\nicefrac{\chi^2}{\textrm{\small dof}}\lesssim 2$. The spread among
the results when different mass cuts are applied is negligible compared to the
statistical error. Thus we use the fit with $m_\pi<500$~MeV to quote a final
result for the scalar radius.

\begin{figure}[h]
\centering
\includegraphics[width=0.5\textwidth]
{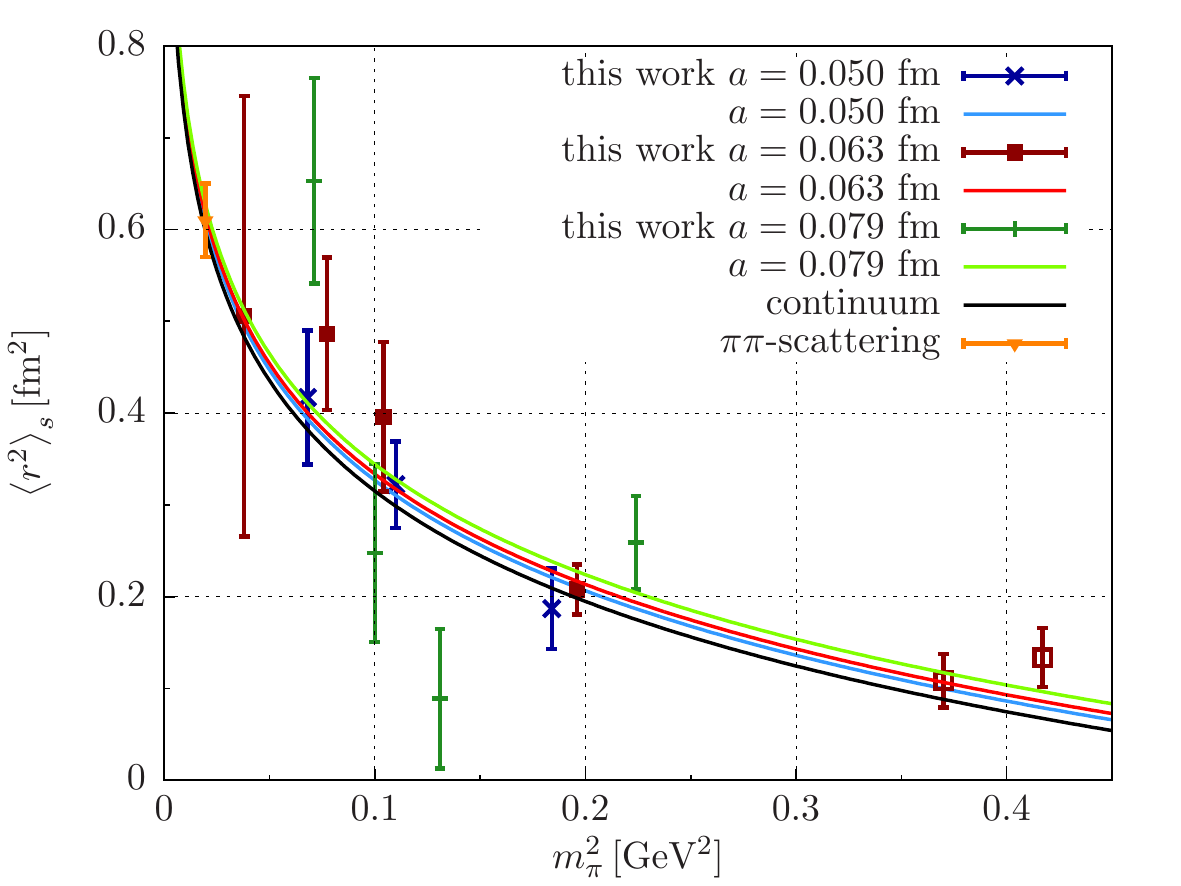}
\caption{Our results for the scalar radius plotted against the
         pion mass $m_\pi^2$ including the result from a combined
         chiral and continuum extrapolation. Different colours denote
         different lattice spacings. The black line is the pion mass
         dependence in the continuum from our combined fit.
         The two right most points (open symbols) at
         $m_\pi^2\approx0.37$~GeV$^2$ (E4)
         and $m_\pi^2\approx0.42$~GeV$^2$ (E3) are not
         included in the fit. The orange point at the physical pion mass
         shows the value from $\pi\pi$-scattering %
         \cite{Colangelo:2001df}.}
\label{fig:combinedchiralcontextra}
\end{figure}

The result of this combined fit to our data is shown in figure
\ref{fig:combinedchiralcontextra}, where the scalar radius is plotted against
the squared pion mass. 
The green, red and blue curves describe the pion mass dependence at individual
values of the lattice spacing, i.e.\ $a=0.079$~fm, $a=0.063$~fm and $0.050$~fm.
Clearly,
the lattice spacing dependence is very small; we find $ b= 4.7\pm12.5$ in our
fit, indicating that a possible dependence on $a^2$ cannot be resolved with the
current statistical accuracy of our data.

The black line in figure \ref{fig:combinedchiralcontextra} shows the pion mass
dependence in the continuum, $a=0$. At the physical pion mass we find for the
scalar radius
\begin{equation}
\left\langle r^2\right\rangle^\pi_{_{\rm S}} =
0.600\pm0.052~\textnormal{fm}^2\,.
\end{equation}
This is in excellent agreement with the value $0.61\pm0.04$~fm$^2$ which was
extracted from $\pi\pi$-scattering in \cite{Colangelo:2001df} (cf. the orange
point in figure \ref{fig:combinedchiralcontextra}). 

One has to note that the pion mass dependence of the scalar radius is
completely driven by $\chi$PT (cf. equation \eqref{eq:scalarrchipt}),
and the fit parameter $\ellf$ only changes the offset of the black curve in
figure \ref{fig:combinedchiralcontextra}. We find that our results for the
radius show the expected pion mass dependence from NLO $\chi$PT, confirming its
applicability.

From the combined fit we can extract the low-energy constant $\ellf$, which
served as a fit parameter. We find
\begin{equation}
 \ellf= 4.54\pm0.30\,,
\end{equation}
which is in very good agreement with the value $\ellf = 4.62\pm0.22$ quoted in
the current FLAG report \cite{Aoki:2013ldr} for the $N_f=2$ flavour theory.

In our previous work we found that the disconnected contribution to the scalar
radius is important, which is in qualitative agreement with what has been found
in partially quenched chiral perturbation theory \cite{Juttner:2011ur}.
We can confirm this using our new data by repeating
the same combined fit for the scalar radius using the connected contribution
only: we indeed find a smaller value for the scalar radius
$\left\langle r^2\right\rangle^{\pi,~\rm conn}_{_{\rm S}}=
0.532\pm0.042~\textnormal{fm}^2$, corresponding to a smaller value for
the low-energy constant $\ellf^{\rm conn}=4.15\pm0.24$, which would
disagree with the determination of $\ellf$ from other processes.

\section{Summary}

We have extended our previous study of the pion scalar radius using
additional lattice spacings and pion masses in order to enable a fully
controlled chiral and continuum extrapolation. We find that discretisation
artefacts are mild, and that the pion mass dependence of the scalar radius
is well described by NLO $\chi$PT. From a combined ex\-tra\-polation to the
physical point, we are able to extract the low-energy constant $\ellf=
4.54\pm0.30$ in good agreement with the FLAG \cite{Aoki:2013ldr} average.

The disconnected part of the pion three-point function contributes significantly
to the slope of the scalar form factor near $Q^2=0$ and thus to the scalar
radius of the pion. The inclusion of the disconnected diagrams is therefore
essential in order to capture the correct physics.

%
%
\begin{acknowledgement}
We acknowledge useful discussions with Andreas J\"uttner, Bastian Brandt and
Harvey B.~Meyer.  Some of our calculations were performed on the ``Wilson'' HPC
Cluster at the Institute for Nuclear Physics, University of Mainz. We thank
Dalibor Djukanovic and Christian Seiwerth for technical support. We are
grateful for computer time allocated to project HMZ21 on the BlueGene
computers ``JUGENE'' and ``JUQUEEN'' at NIC, J\"ulich. This research has been
supported in part by the DFG in the SFB~1044. We are grateful to our
colleagues in the CLS initiative for sharing ensembles.
\end{acknowledgement}
%
%

%
%
\bibliographystyle{h-physrev4}
\bibliography{scalarformfactor}

\end{document}